\documentclass[aip, apl,amsmath,amssymb,reprint,floatfix]{revtex4-1}
\usepackage{graphicx}
\usepackage{dcolumn}
\usepackage{bm}
\usepackage{textcomp}
\usepackage[utf8]{inputenc}

\begin{document}

\title{Turn-key module for neutron scattering with sub-micro-eV resolution}

\author{R. Georgii}
\email{Robert.Georgii@frm2.tum.de}
\affiliation{Forschungsneutronenquelle Heinz-Maier Leibnitz,
  Technische Universit\"at M\"unchen, 85748 Garching, Germany} 
\affiliation{Physik-Department E21,
  Technische Universit\"at M\"unchen, 85748 Garching, Germany}

\author{G. Brandl}
\affiliation{Physik-Department E21,
  Technische Universit\"at M\"unchen, 85748 Garching, Germany}
 \affiliation{Forschungsneutronenquelle Heinz-Maier Leibnitz,
  Technische Universit\"at M\"unchen, 85748 Garching, Germany} 

\author{N. Arend}
\affiliation{Forschungszentrum Jülich GmbH, JCNS-1 \& ICS-1, Oak Ridge National Laboratory, Oak Ridge, TN 37831, USA} 
\affiliation{Physik-Department E21,
  Technische Universit\"at M\"unchen, 85748 Garching, Germany}

\author{W. H\"au{\ss}ler}
\affiliation{Forschungsneutronenquelle Heinz-Maier Leibnitz,
  Technische Universit\"at M\"unchen, 85748 Garching, Germany} 
\affiliation{Physik-Department E21,
  Technische Universit\"at M\"unchen, 85748 Garching, Germany}

\author{A. Tischendorf}
\affiliation{Physik-Department E21,
  Technische Universit\"at M\"unchen, 85748 Garching, Germany}

\author{C. Pfleiderer}
\affiliation{Physik-Department E21,
  Technische Universit\"at M\"unchen, 85748 Garching, Germany}

\author{P. B\"oni}
\affiliation{Physik-Department E21,
  Technische Universit\"at M\"unchen, 85748 Garching, Germany}

\author{J. Lal}
\affiliation{Materials Science Division, Argonne National Lab, Argonne IL-60439, USA}

\date{\today}

\begin{abstract}
  We report the development of a compact turn-key module that boosts the
  resolution in quasi-elastic neutron scattering by several orders of magnitude
  down to the low sub-\textmu eV range. It is based on a pair of neutron
  resonance spin flippers that generate a well defined temporal intensity
  modulation, also known as MIEZE (Modulation of IntEnsity by Zero Effort). The
  module may be used under versatile conditions, in particular in applied
  magnetic fields and for depolarising and incoherently scattering  samples. We demonstrate the power of
  MIEZE in studies of the helimagnetic order in MnSi under applied magnetic
  fields. 
   \newline
    
  \noindent \copyright (2011) American Institute of Physics. This article may be downloaded for personal use only. Any other use requires
   prior permission of the author and the American Institute of Physics. 
   
   \noindent The following article appeared in Applied Physics Letters and may be found at URL: http://link.aip.org/link/?apl/98/073505 
\end{abstract}
\maketitle

Neutron scattering is an extremely powerful technique for studies of the
dynamical properties of condensed matter systems. Prominent examples of great
current interest concern the spin dynamics in transition metal and rare earth
compounds and diffusive processes in soft matter systems such as proteins,
liquid crystals and emulsions. A precondition to unravel some of the most
important scientific challenges is the need for high energy and momentum
resolution.

Conventional neutron scattering techniques such as triple-axis and
time-of-flight spectroscopy provide momentum resolved energy resolutions of the
order of $\simeq$ 10 \textmu eV.  Backscattering reaches the sub-\textmu eV
regime, however sacrificing momentum resolution.  This is contrasted by neutron
spin-echo (NSE) methods \cite{Mezei:72, Golub:87(M)}, which offer high energy
and momentum resolutions in the low sub-\textmu eV range -- several orders of
magnitude below the typical resolutions of conventional techniques. However,
because NSE scattering uses polarised neutrons it is inherently sensitive to the
depolarisation of the neutron beam. Therefore, it is technically very demanding
to perform NSE measurements under applied magnetic fields or in depolarising
samples\cite{Pappas:08, Pappas:09} such as superconductors,
ferromagnets\cite{Mezei} or protonated soft matter systems.

\begin{figure}[b]
  \includegraphics[width=\linewidth]{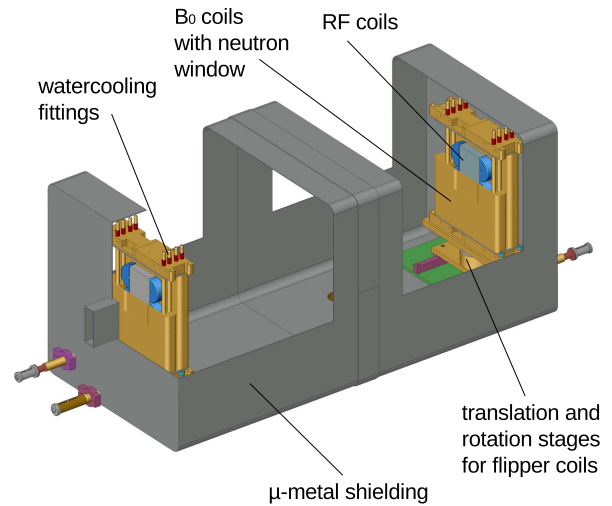}
  \caption{\label{Fig:Box}
    The ``MIEZE box'' as used for the MIEZE setup at MIRA.}
\end{figure}

In this Letter we report the development of a turn-key module, the MIEZE box
shown in Fig.\,\ref{Fig:Box}.  In combination with a polariser, a polarising
analyser and a fast detector this box allows to improve the energy resolution in
all types of neutron scattering instruments capable of studying quasi-elastic
scattering, notably diffractometers, small-angle neutron scattering cameras and
reflectometers (especially instruments for small $q$ and cold neutrons), down to
the sub-\textmu eV range.  { The module is based on the so-called MIEZE-I
  technique (Modulation of IntEnsity by Zero Effort, type I) \cite{Gaehler:92,
    Koeppe:96, Besenboeck:98, Keller:02, Bleuel:06, Kawabata:06}, where the
  modulation of the beam is performed upstream of the sample.  Therefore, in
  contrast to NSE, the MIEZE module we describe may even be used under
  depolarising conditions in or around the sample. }This routinely allows
neutron scattering studies with the highest possible energy resolution in a wide
range of materials.

\begin{figure}
  \includegraphics[width=\linewidth]{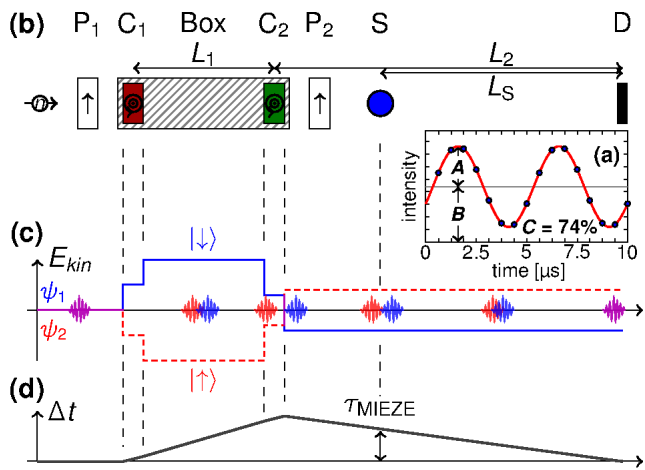}
  \caption{\label{figure1} {\bf (a)} A typical MIEZE signal at the detector
    position (see text for details). {\bf (b)}~Schematic of a complete MIEZE
    setup, showing the polariser (P$_1$), the zero field of the MIEZE box
    (hatched) with two $\pi$-flipper coils (C$_1$, C$_2$), the analyser (P$_2$),
    the sample (S) and the detector (D). {\bf (c)}~Kinetic energy splitting for
    the spin-down ($\psi_1$) and spin-up ($\psi_2$) states of the neutrons along
    the flight path due to the $\pi$-flipper coils. {\bf (d)}~Temporal delay
    $\Delta t$ of the spin states along the flight path.  The splitting reaches
    its maximum after the second flipper coil and vanishes at the detector
    position.
  }
\end{figure}

Qualitatively, the MIEZE-I technique is based on a harmonic intensity modulation
of the neutron beam, where the contrast $C$ is given by the ratio of the
amplitude $A$ to the average signal $B$ as shown in Fig. \ref{figure1}a.  Using
a phase-locked pair of two resonance spin flippers\cite{Keller:02, Arend:06},
which operate at slightly different frequencies, $\nu_1$ and $\nu_2$, induces a
slow rotation of the polarisation direction of the neutrons, which is
subsequently converted into an intensity beating by means of a polarising
analyser.

{ While former experiments with the MIEZE-I technique have been successful,
  we managed to implement MIEZE-I as a routine technique through a redesign of
  the neutron resonance spin flippers.  Instead of wire-wound B$_0$ coils we use
  electro-erosion machined coil windings where a better
  definition of the magnetic field boundaries is obtained. They consist of a specially selected Al alloy
  with much less small angle scattering and a higher transmission\cite{Arend:06}. 
  Together with a more reproducible mounting  of the RF coils this
  results in stable $\pi$-flips over a wide range of RF frequencies with low
  small angle scattering background. }

A description of the MIEZE-I principle pointing out its vicinity to TOF methods
is illustrated in Fig.\,\ref{figure1}; for a proper quantum mechanical
description we refer to the literature (cf Ref.\,\cite{Arend:04}).  When a
neutron arrives at the first spin flipper with its polarisation perpendicular to
the static field in the flipper, the correlation volumes (or wave-packages)
corresponding to the spin-up and spin-down spin states are prepared (cf
Fig.\,\ref{figure1}\,(c)).  While the kinetic energy of the spin-down state
increases, the kinetic energy of the spin-up state decreases.  Therefore, the
correlation volumes for the spin-down and spin-up states arrive at different
times at the second resonance spin flipper placed at a distance $L_1$ behind the
first spin flipper.  This second spin flipper inverts the energy splitting of
the spin states, reducing the kinetic energy of the spin-down state and
increasing the kinetic energy of the spin-up state. Therefore the correlation
volumes, overlap again at a distance $L_2$ behind the second spin flipper, given
by $L_2 = L_1\,(\nu_2 / \nu_1 - 1)^{-1}$ where $\nu_2>\nu_1$.

An analyser at an arbitrary position between the second spin flipper and the
detector (the latter is located where the correlation volumes meet) projects out
the intensity of the interference pattern of the spin-up and spin-down states.
As the correlation volumes of the spin-up and spin-down states have different
energies, the interference pattern exhibits the intensity modulation of contrast
$C$ referred to above.

To explain how this intensity modulation may be exploited in experimental
studies using two single neutron resonance spin flippers, we show in
Fig.\,\ref{figure1}\,(d) the delay between the two correlation volumes,
$\Delta t$.  The correlation volumes probe the sample at different times with
the delay given by \cite{Comment}
$\tau_{\rm MIEZE} = 2\cdot\hbar(2\pi\Delta\nu) L_S/(mv^3)$, where
2$\Delta\nu = 2(\nu_2 - \nu_1) $ is the frequency of the resulting MIEZE signal,
$L_S$ is the distance between sample and detector, and $m$ and $v$ are the mass
and average velocity of the neutrons.  By overlapping these volumes at the
detector, one obtains a signal contrast $C$ which is directly proportional to
the intermediate scattering function $S(q,\tau_{\rm MIEZE})$, i.e., the
information on the dynamics on this time scale.  Further, for quasi-elastic
scattering with an assumed Lorentzian line shape with half-width $\Gamma$, the
normalised intermediate scattering function is given by \cite{Keller:02}
$ S(q, \tau)/S(q, 0) = \exp\left(-\Gamma(q) \tau \right)$, where $S(q,0)$
corresponds to the intermediate scattering function of a purely elastically
scattering sample.

\begin{figure}
\includegraphics[width=0.8\linewidth]{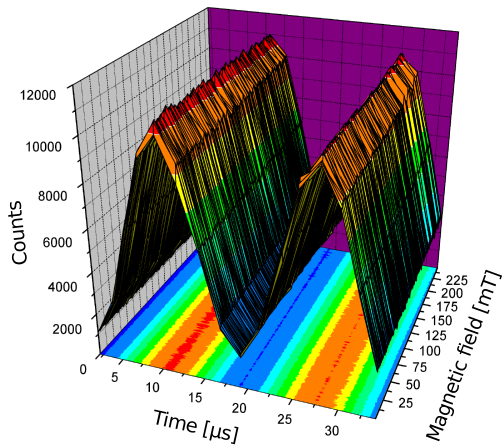}
\caption{\label{Fig:Magnet} {The MIEZE signal for the direct beam and a magnetic field
  produced with an electromagnet \cite{Arend:06}. Note that the contrast does not
  depend on the field strength.}}
\end{figure}

The MIEZE-I technique is similar to conventional neutron resonance spin echo
(NRSE) methods, where the two spin flippers before the sample correspond to the
first arm of an NRSE instrument. Moreover, the MIEZE time $\tau_{\rm MIEZE}$ is
equivalent to the spin echo time in NSE and NRSE instruments \cite{Keller:02}.
However, there is a distinct difference between MIEZE-I and NSE/NRSE.  Placing
the polarising analyser behind the second spin flipper and before the sample,
the MIEZE-I technique becomes insensitive to effects of the sample or sample
environment on the polarisation of the neutron beam, i.e. for example
depolarisation or applied magnetic fields. { This is demonstrated in Fig.\,\ref{Fig:Magnet},
where the contrast of the direct beam is plotted versus
a magnetic field up to 0.2 T and no effect on the signal contrast can be observed.}

The MIEZE experiments were carried out at the diffractometer MIRA at FRM II using
neutrons with a wavelength $\lambda=10.4\,$\AA$\pm5\%$, i.e., a mean velocity
$v=380\,{\rm m/s}$.  The first and second spin flipper { being 0.9 m apart} were operating at
frequencies in the range $46\,{\rm kHz}<\nu_1<200\,{\rm kHz}$ and
$69\,{\rm kHz}<\nu_2<300\,{\rm kHz}$, respectively, providing a beating
frequency in the range $46\,{\rm kHz}<2\Delta\nu<200\,{\rm kHz}$.   The distance
between the sample and detector  was $860\,{\rm mm}$.  Taken together MIEZE times could be accessed in a range
$280\,{\rm ps}<\tau_{\rm MIEZE}<1230\,{\rm ps}$.  A 0.3 mm thick $^6$Li doped
glass scintillator with a photomultiplier was used as a fast detector.

We note that the temporal and thus spatial separation of the spin-up and
spin-down states makes MIEZE-I sensitive to path length differences between the
first RF-flipper and the detector, e.g.\ due to the large divergence of the
beam, the finite size of the sample or the finite thickness of the neutron
detector.  { However, for the wavelength of 10.4\,\AA, frequencies in the
  range 46\,kHz to 200\,kHz for the MIEZE signal, typical sample sizes of 10 mm,
  and scattering angles of the order of a few degrees we measured with a
  standard sample used for resolution measurements a contrast reduction of less than 10\% up
  to $q =0.05$ \AA$^{-1}$ being in agreement with the results of calculations of the path length and precession phase differences.}

\begin{figure}
\includegraphics[width=0.8\linewidth]{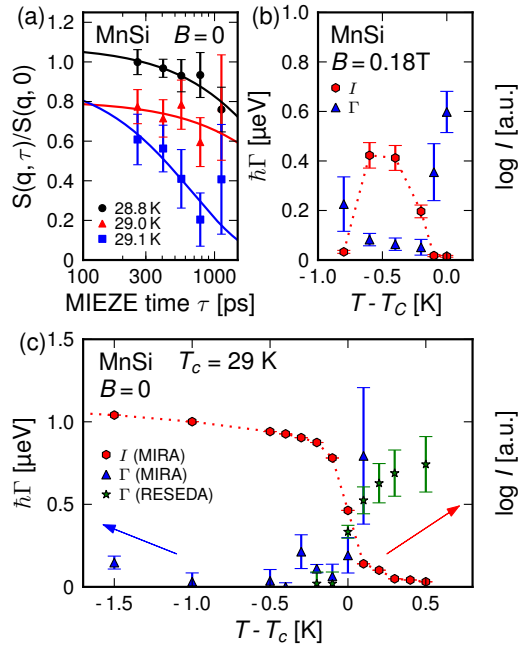}
\caption{\label{figure3} {\bf (a)} Typical normalised intermediate scattering functions
  $S(q, \tau)$ in the helimagnetic state of MnSi at selected temperatures. 
     {\bf (b)}{ Line width $\Gamma$ in the A-phase of MnSi at $B = 0.18$ T. The total scattering
  intensity is shown as solid hexagons, MIEZE data as solid triangles. 
  The data are normalised to the line width of the helical phase at $T = 3$ K.
  {\bf(c)} Line width $\Gamma$ of the magnetic order in MnSi at $B=0$. The data
   are normalised to the line width at $T = 3$ K. NRSE  data
  from the instrument RESEDA are shown as stars.}}
\end{figure}

To demonstrate the power of MIEZE-I in studies of field-induced forms of
magnetic order normally not accessible to NSE/NRSE, we have investigated the
cubic B20 compound MnSi \cite{Ishikawa:76}.  MnSi orders helimagnetically below
a transition temperature, in our case at $T_c=29.0\,{\rm K}$.  { The large
  pitch of the helix $\Lambda\approx180\,$\AA{} implies that the associated
  magnetic Bragg peak at $q = 0.035$ \AA$^{-1}$ can be accessed in a forward
  scattering configuration.}  Just below $T_c$ at $0.15\,{\rm T}<B<0.2\,{\rm
  T}$,
the A-phase is observed. Here a special form of magnetic order, a skyrmion
lattice, is observed\cite{Muehlbauer:09b}.

Shown in Fig.\,\ref{figure3}\,(a) are typical data in the helimagnetic state
($B=0$) of the normalised intermediate scattering function $S(q, \tau)/S(q, 0)$
for various temperatures.  Data were normalised with respect to resolution
measurements in the helical ordered sample at low temperatures, where the
magnetic structure is supposed to be static.  The solid lines are the result
of fits to exponential functions.
The presence of a second process on a shorter timescale as indicated
by the fact that the fits are not converging to $1$ for $\tau \rightarrow 0$
around T$_c$ is still under discussion and will be published elsewhere.
The resulting line widths (solid triangels) as a function of temperature are
shown in Fig.\,\ref{figure3}\,(c) (solid hexagones show the elastic magnetic
intensity). While the magnetic order is resolution limited below $T_c$, there is
broadening above $T_c$.  { For comparison, NRSE data from the same system are
  shown as stars. 
 
 The measured $\Gamma$  in the A-Phase of MnSi at $B = 0.18$
  T (shown in Fig.\,\ref{figure3}\,(b)) is similar to the one in the helical phase.  This demonstrates
  that even under applied magnetic fields the MIEZE-I technique may be readily
  used.}

We have recently used our MIEZE box at the beam line CG-1D at HFIR at Oak Ridge
National Laboratory \cite{HFIR}.  Here and on MIRA to set up and adjust the
MIEZE box requires less than a few hours making it indeed a turn-key measurement
option.

In conclusion we have developed a compact turn-key MIEZE module, that allows to
improve the energy resolution of various neutron scattering instruments used for
quasi-elastic scattering into the sub-\textmu eV range 
{ for typical sample sizes of 1 cm in diameter and $q$ values up to 0.05
  \AA$^{-1}$}.  We have demonstrated the power of this technique by small angle scattering  studies of the magnetic order of MnSi in applied magnetic fields.  One particular strength of the MIEZE-I technique is its
insensitivity to depolarising conditions at the sample position, which is
not easily achievable using standard spin echo techniques. This illustrates the wide range of
scientific challenges that may now be addressed with sub-\textmu eV resolution.

We gratefully acknowledge support by R. Schwikowski, A. Mantwill, M. Wipp and
the team of FRM~II.  Many thanks to R. G\"ahler, L. Robertson, I. Anderson and
W. Petry for helpful discussions and support.  Financial support through the
German Science Foundation (DFG), the Oak Ridge National Laboratory and 
the European Commission under the 7th Framework Program, Contract
no. CP-CSA INFRA-2008-1.1.1 Number 226507-NMI3 is acknowledged.


\begin{thebibliography}{17}%
\makeatletter
\providecommand \@ifxundefined [1]{%
 \@ifx{#1\undefined}
}%
\providecommand \@ifnum [1]{%
 \ifnum #1\expandafter \@firstoftwo
 \else \expandafter \@secondoftwo
 \fi
}%
\providecommand \@ifx [1]{%
 \ifx #1\expandafter \@firstoftwo
 \else \expandafter \@secondoftwo
 \fi
}%
\providecommand \natexlab [1]{#1}%
\providecommand \enquote  [1]{``#1''}%
\providecommand \bibnamefont  [1]{#1}%
\providecommand \bibfnamefont [1]{#1}%
\providecommand \citenamefont [1]{#1}%
\providecommand \href@noop [0]{\@secondoftwo}%
\providecommand \href [0]{\begingroup \@sanitize@url \@href}%
\providecommand \@href[1]{\@@startlink{#1}\@@href}%
\providecommand \@@href[1]{\endgroup#1\@@endlink}%
\providecommand \@sanitize@url [0]{\catcode `\\12\catcode `\$12\catcode
  `\&12\catcode `\#12\catcode `\^12\catcode `\_12\catcode `\%12\relax}%
\providecommand \@@startlink[1]{}%
\providecommand \@@endlink[0]{}%
\providecommand \url  [0]{\begingroup\@sanitize@url \@url }%
\providecommand \@url [1]{\endgroup\@href {#1}{\urlprefix }}%
\providecommand \urlprefix  [0]{URL }%
\providecommand \Eprint [0]{\href }%
\@ifxundefined \urlstyle {%
  \providecommand \doi  [0]{\begingroup \@sanitize@url \@doi}%
  \providecommand \@doi [1]{\endgroup \@@startlink {\doibase
  #1}doi:\discretionary {}{}{}#1\@@endlink }%
}{%
  \providecommand \doi  [0]{doi:\discretionary{}{}{}\begingroup
  \urlstyle{rm}\Url }%
}%
\providecommand \doibase [0]{http://dx.doi.org/}%
\providecommand \Doi [0]{\begingroup \@sanitize@url \@Doi }%
\providecommand \@Doi  [1]{\endgroup\@@startlink{\doibase#1}\@@Doi}%
\providecommand \@@Doi [1]{#1\@@endlink}%
\providecommand \selectlanguage [0]{\@gobble}%
\providecommand \bibinfo  [0]{\@secondoftwo}%
\providecommand \bibfield  [0]{\@secondoftwo}%
\providecommand \translation [1]{[#1]}%
\providecommand \BibitemOpen [0]{}%
\providecommand \bibitemStop [0]{}%
\providecommand \bibitemNoStop [0]{.\EOS\space}%
\providecommand \EOS [0]{\spacefactor3000\relax}%
\providecommand \BibitemShut  [1]{\csname bibitem#1\endcsname}%
\bibitem [{\citenamefont {Mezei}(1972)}]{Mezei:72}%
  \BibitemOpen
  \bibfield  {author} {\bibinfo {author} {\bibfnamefont {F.}~\bibnamefont
  {Mezei}},\ }\Doi {DOI: 10.1007/BF01394523} {\bibfield  {journal} {\bibinfo
  {journal} {Z. Phys. A},\ }\textbf {\bibinfo {volume} {255}},\ \bibinfo
  {pages} {146} (\bibinfo {year} {1972})}\BibitemShut {NoStop}%
\bibitem [{\citenamefont {Golub}\ and\ \citenamefont
  {Gähler}(1987)}]{Golub:87(M)}%
  \BibitemOpen
  \bibfield  {author} {\bibinfo {author} {\bibfnamefont {R.}~\bibnamefont
  {Golub}}\ and\ \bibinfo {author} {\bibfnamefont {R.}~\bibnamefont
  {Gähler}},\ }\Doi {DOI: 10.1016/0375-9601(87)90760-2} {\bibfield  {journal}
  {\bibinfo  {journal} {Phys. Lett. A},\ }\textbf {\bibinfo {volume} {123}},\
  \bibinfo {pages} {43} (\bibinfo {year} {1987})}\BibitemShut {NoStop}%
\bibitem [{\citenamefont {Pappas}\ \emph {et~al.}(2008)\citenamefont {Pappas},
  \citenamefont {Lelievre-Berna}, \citenamefont {Bentley}, \citenamefont
  {Bourgeat-Lami}, \citenamefont {Moskvin}, \citenamefont {Thomas},
  \citenamefont {Grigoriev},\ and\ \citenamefont {Dyadkin}}]{Pappas:08}%
  \BibitemOpen
  \bibfield  {author} {\bibinfo {author} {\bibfnamefont {C.}~\bibnamefont
  {Pappas}}, \bibinfo {author} {\bibfnamefont {E.}~\bibnamefont
  {Lelievre-Berna}}, \bibinfo {author} {\bibfnamefont {P.}~\bibnamefont
  {Bentley}}, \bibinfo {author} {\bibfnamefont {E.}~\bibnamefont
  {Bourgeat-Lami}}, \bibinfo {author} {\bibfnamefont {E.}~\bibnamefont
  {Moskvin}}, \bibinfo {author} {\bibfnamefont {M.}~\bibnamefont {Thomas}},
  \bibinfo {author} {\bibfnamefont {S.}~\bibnamefont {Grigoriev}}, \ and\
  \bibinfo {author} {\bibfnamefont {V.}~\bibnamefont {Dyadkin}},\ }\Doi {DOI:
  10.1016/j.nima.2008.04.078} {\bibfield  {journal} {\bibinfo  {journal} {NIM
  A},\ }\textbf {\bibinfo {volume} {592}},\ \bibinfo {pages} {420 } (\bibinfo
  {year} {2008})}\BibitemShut {NoStop}%
\bibitem [{\citenamefont {Pappas}\ \emph {et~al.}(2009)\citenamefont {Pappas},
  \citenamefont {Lelievre-Berna}, \citenamefont {Falus}, \citenamefont
  {Bentley}, \citenamefont {Moskvin}, \citenamefont {Grigoriev}, \citenamefont
  {Fouquet},\ and\ \citenamefont {Farago}}]{Pappas:09}%
  \BibitemOpen
  \bibfield  {author} {\bibinfo {author} {\bibfnamefont {C.}~\bibnamefont
  {Pappas}}, \bibinfo {author} {\bibfnamefont {E.}~\bibnamefont
  {Lelievre-Berna}}, \bibinfo {author} {\bibfnamefont {P.}~\bibnamefont
  {Falus}}, \bibinfo {author} {\bibfnamefont {P.~M.}\ \bibnamefont {Bentley}},
  \bibinfo {author} {\bibfnamefont {E.}~\bibnamefont {Moskvin}}, \bibinfo
  {author} {\bibfnamefont {S.}~\bibnamefont {Grigoriev}}, \bibinfo {author}
  {\bibfnamefont {P.}~\bibnamefont {Fouquet}}, \ and\ \bibinfo {author}
  {\bibfnamefont {B.}~\bibnamefont {Farago}},\ }\Doi
  {10.1103/PhysRevLett.102.197202} {\bibfield  {journal} {\bibinfo  {journal}
  {Phys. Rev. Lett.},\ }\textbf {\bibinfo {volume} {102}},\ \bibinfo {pages}
  {197202} (\bibinfo {year} {2009})}\BibitemShut {NoStop}%
\bibitem [{\citenamefont {Farago}\ and\ \citenamefont {Mezei}(1986)}]{Mezei}%
  \BibitemOpen
  \bibfield  {author} {\bibinfo {author} {\bibfnamefont {B.}~\bibnamefont
  {Farago}}\ and\ \bibinfo {author} {\bibfnamefont {F.}~\bibnamefont {Mezei}},\
  }\href@noop {} {\bibfield  {journal} {\bibinfo  {journal} {Physica B},\
  }\textbf {\bibinfo {volume} {136}},\ \bibinfo {pages} {100} (\bibinfo {year}
  {1986})}\BibitemShut {NoStop}%
\bibitem [{\citenamefont {Gähler}\ \emph {et~al.}(1992)\citenamefont
  {Gähler}, \citenamefont {Golub},\ and\ \citenamefont {Keller}}]{Gaehler:92}%
  \BibitemOpen
  \bibfield  {author} {\bibinfo {author} {\bibfnamefont {R.}~\bibnamefont
  {Gähler}}, \bibinfo {author} {\bibfnamefont {R.}~\bibnamefont {Golub}}, \
  and\ \bibinfo {author} {\bibfnamefont {T.}~\bibnamefont {Keller}},\ }\Doi
  {DOI: 10.1016/0921-4526(92)90503-K} {\bibfield  {journal} {\bibinfo
  {journal} {Physica B},\ }\textbf {\bibinfo {volume} {180-181}},\ \bibinfo
  {pages} {899} (\bibinfo {year} {1992})}\BibitemShut {NoStop}%
\bibitem [{\citenamefont {Köppe}\ \emph {et~al.}(1996)\citenamefont {Köppe},
  \citenamefont {Hank}, \citenamefont {Wuttke}, \citenamefont {Petry},
  \citenamefont {Gähler},\ and\ \citenamefont {Kahn}}]{Koeppe:96}%
  \BibitemOpen
  \bibfield  {author} {\bibinfo {author} {\bibfnamefont {M.}~\bibnamefont
  {Köppe}}, \bibinfo {author} {\bibfnamefont {P.}~\bibnamefont {Hank}},
  \bibinfo {author} {\bibfnamefont {J.}~\bibnamefont {Wuttke}}, \bibinfo
  {author} {\bibfnamefont {W.}~\bibnamefont {Petry}}, \bibinfo {author}
  {\bibfnamefont {R.}~\bibnamefont {Gähler}}, \ and\ \bibinfo {author}
  {\bibfnamefont {R.}~\bibnamefont {Kahn}},\ }\Doi {DOI:
  10.1080/10238169608200092} {\bibfield  {journal} {\bibinfo  {journal} {J.
  Neutron Res.},\ }\textbf {\bibinfo {volume} {4}},\ \bibinfo {pages} {261}
  (\bibinfo {year} {1996})}\BibitemShut {NoStop}%
\bibitem [{\citenamefont {Besenböck}\ \emph {et~al.}(1998)\citenamefont
  {Besenböck}, \citenamefont {Gähler}, \citenamefont {Hank}, \citenamefont
  {Kahn}, \citenamefont {Köppe}, \citenamefont {Novion}, \citenamefont
  {Petry},\ and\ \citenamefont {Wuttke}}]{Besenboeck:98}%
  \BibitemOpen
  \bibfield  {author} {\bibinfo {author} {\bibfnamefont {W.}~\bibnamefont
  {Besenböck}}, \bibinfo {author} {\bibfnamefont {R.}~\bibnamefont {Gähler}},
  \bibinfo {author} {\bibfnamefont {P.}~\bibnamefont {Hank}}, \bibinfo {author}
  {\bibfnamefont {R.}~\bibnamefont {Kahn}}, \bibinfo {author} {\bibfnamefont
  {M.}~\bibnamefont {Köppe}}, \bibinfo {author} {\bibfnamefont {C.~H.~D.}\
  \bibnamefont {Novion}}, \bibinfo {author} {\bibfnamefont {W.}~\bibnamefont
  {Petry}}, \ and\ \bibinfo {author} {\bibfnamefont {J.}~\bibnamefont
  {Wuttke}},\ }\Doi {DOI: 10.1080/10238169808200231} {\bibfield  {journal}
  {\bibinfo  {journal} {J. Neutron Res.},\ }\textbf {\bibinfo {volume} {7}},\
  \bibinfo {pages} {65} (\bibinfo {year} {1998})}\BibitemShut {NoStop}%
\bibitem [{\citenamefont {Keller}\ \emph {et~al.}(2002)\citenamefont {Keller},
  \citenamefont {Golub},\ and\ \citenamefont {Gähler}}]{Keller:02}%
  \BibitemOpen
  \bibfield  {author} {\bibinfo {author} {\bibfnamefont {T.}~\bibnamefont
  {Keller}}, \bibinfo {author} {\bibfnamefont {R.}~\bibnamefont {Golub}}, \
  and\ \bibinfo {author} {\bibfnamefont {R.}~\bibnamefont {Gähler}},\ }in\
  \href@noop {} {\emph {\bibinfo {booktitle} {Scattering and Inverse Scattering
  in Pure and Applied Science}}}\ (\bibinfo  {publisher} {Academic Press},\
  \bibinfo {address} {London},\ \bibinfo {year} {2002})\ pp.\ \bibinfo {pages}
  {1264--1286}\BibitemShut {NoStop}%
\bibitem [{\citenamefont {Bleuel}\ \emph {et~al.}(2006)\citenamefont {Bleuel},
  \citenamefont {Bröll}, \citenamefont {Lang}, \citenamefont {Littrell},
  \citenamefont {Gähler},\ and\ \citenamefont {Lal}}]{Bleuel:06}%
  \BibitemOpen
  \bibfield  {author} {\bibinfo {author} {\bibfnamefont {M.}~\bibnamefont
  {Bleuel}}, \bibinfo {author} {\bibfnamefont {M.}~\bibnamefont {Bröll}},
  \bibinfo {author} {\bibfnamefont {E.}~\bibnamefont {Lang}}, \bibinfo {author}
  {\bibfnamefont {K.}~\bibnamefont {Littrell}}, \bibinfo {author}
  {\bibfnamefont {R.}~\bibnamefont {Gähler}}, \ and\ \bibinfo {author}
  {\bibfnamefont {J.}~\bibnamefont {Lal}},\ }\Doi {DOI:
  10.1016/j.physb.2005.10.124} {\bibfield  {journal} {\bibinfo  {journal}
  {Physica B},\ }\textbf {\bibinfo {volume} {371}},\ \bibinfo {pages} {297}
  (\bibinfo {year} {2006})}\BibitemShut {NoStop}%
\bibitem [{\citenamefont {Kawabata}\ \emph {et~al.}(2006)\citenamefont
  {Kawabata}, \citenamefont {Hino}, \citenamefont {Kitaguchi}, \citenamefont
  {Hayashida}, \citenamefont {Tasaki}, \citenamefont {Ebisawa}, \citenamefont
  {Yamazaki}, \citenamefont {Maruyama}, \citenamefont {Seto}, \citenamefont
  {Nagao},\ and\ \citenamefont {Kanaya}}]{Kawabata:06}%
  \BibitemOpen
  \bibfield  {author} {\bibinfo {author} {\bibfnamefont {Y.}~\bibnamefont
  {Kawabata}}, \bibinfo {author} {\bibfnamefont {M.}~\bibnamefont {Hino}},
  \bibinfo {author} {\bibfnamefont {M.}~\bibnamefont {Kitaguchi}}, \bibinfo
  {author} {\bibfnamefont {H.}~\bibnamefont {Hayashida}}, \bibinfo {author}
  {\bibfnamefont {S.}~\bibnamefont {Tasaki}}, \bibinfo {author} {\bibfnamefont
  {T.}~\bibnamefont {Ebisawa}}, \bibinfo {author} {\bibfnamefont
  {D.}~\bibnamefont {Yamazaki}}, \bibinfo {author} {\bibfnamefont
  {R.}~\bibnamefont {Maruyama}}, \bibinfo {author} {\bibfnamefont
  {H.}~\bibnamefont {Seto}}, \bibinfo {author} {\bibfnamefont {M.}~\bibnamefont
  {Nagao}}, \ and\ \bibinfo {author} {\bibfnamefont {T.}~\bibnamefont
  {Kanaya}},\ }\Doi {DOI: 10.1016/j.physb.2006.05.387} {\bibfield  {journal}
  {\bibinfo  {journal} {Physica B},\ }\textbf {\bibinfo {volume} {385-386}},\
  \bibinfo {pages} {1122} (\bibinfo {year} {2006})}\BibitemShut {NoStop}%
\bibitem [{\citenamefont {Brandl}\ \emph {et~al.}()\citenamefont {Brandl},
  \citenamefont {Bleuel}, \citenamefont {Robertson}, \citenamefont {Crow},
  \citenamefont {Lal},\ and\ \citenamefont {Georgii}}]{HFIR}%
  \BibitemOpen
  \bibfield  {author} {\bibinfo {author} {\bibfnamefont {G.}~\bibnamefont
  {Brandl}}, \bibinfo {author} {\bibfnamefont {M.}~\bibnamefont {Bleuel}},
  \bibinfo {author} {\bibfnamefont {L.}~\bibnamefont {Robertson}}, \bibinfo
  {author} {\bibfnamefont {L.}~\bibnamefont {Crow}}, \bibinfo {author}
  {\bibfnamefont {J.}~\bibnamefont {Lal}}, \ and\ \bibinfo {author}
  {\bibfnamefont {R.}~\bibnamefont {Georgii}},\ }\href@noop {} {}\bibinfo
  {note} {Unpublished}\BibitemShut {NoStop}%
\bibitem [{\citenamefont {Arend}(2007)}]{Arend:06}%
  \BibitemOpen
  \bibfield  {author} {\bibinfo {author} {\bibfnamefont {N.}~\bibnamefont
  {Arend}},\ }\href@noop {} {Ph.D. thesis},\ \bibinfo  {school} {Technical
  University of Munich} (\bibinfo {year} {2007})\BibitemShut {NoStop}%
\bibitem [{\citenamefont {Arend}\ \emph {et~al.}(2004)\citenamefont {Arend},
  \citenamefont {Gähler}, \citenamefont {Keller}, \citenamefont {Georgii},
  \citenamefont {Hils},\ and\ \citenamefont {Böni}}]{Arend:04}%
  \BibitemOpen
  \bibfield  {author} {\bibinfo {author} {\bibfnamefont {N.}~\bibnamefont
  {Arend}}, \bibinfo {author} {\bibfnamefont {R.}~\bibnamefont {Gähler}},
  \bibinfo {author} {\bibfnamefont {T.}~\bibnamefont {Keller}}, \bibinfo
  {author} {\bibfnamefont {R.}~\bibnamefont {Georgii}}, \bibinfo {author}
  {\bibfnamefont {T.}~\bibnamefont {Hils}}, \ and\ \bibinfo {author}
  {\bibfnamefont {P.}~\bibnamefont {Böni}},\ }\Doi
  {doi:10.1016/j.physleta.2004.04.062} {\bibfield  {journal} {\bibinfo
  {journal} {Phys. Lett. A},\ }\textbf {\bibinfo {volume} {327}},\ \bibinfo
  {pages} {21} (\bibinfo {year} {2004})}\BibitemShut {NoStop}%
\bibitem [{Com()}]{Comment}%
  \BibitemOpen
  \href@noop {} {}\bibinfo {note} {{The pre-factor 2 is inherent to the
  NRSE/MIEZE technique when using the resonance flippers in the non-bootstrap
  mode.}}\BibitemShut {Stop}%
\bibitem [{\citenamefont {Ishikawa}\ \emph {et~al.}(1976)\citenamefont
  {Ishikawa}, \citenamefont {Tajima}, \citenamefont {Bloch},\ and\
  \citenamefont {Roth}}]{Ishikawa:76}%
  \BibitemOpen
  \bibfield  {author} {\bibinfo {author} {\bibfnamefont {Y.}~\bibnamefont
  {Ishikawa}}, \bibinfo {author} {\bibfnamefont {K.}~\bibnamefont {Tajima}},
  \bibinfo {author} {\bibfnamefont {D.}~\bibnamefont {Bloch}}, \ and\ \bibinfo
  {author} {\bibfnamefont {M.}~\bibnamefont {Roth}},\ }\href@noop {} {\bibfield
   {journal} {\bibinfo  {journal} {Solid State Commun.},\ }\textbf {\bibinfo
  {volume} {19}},\ \bibinfo {pages} {525} (\bibinfo {year} {1976})}\BibitemShut
  {NoStop}%
\bibitem [{\citenamefont {Mühlbauer}\ \emph {et~al.}(2009)\citenamefont
  {Mühlbauer}, \citenamefont {Binz}, \citenamefont {Jonietz}, \citenamefont
  {Pfleiderer}, \citenamefont {Rosch}, \citenamefont {Neubauer}, \citenamefont
  {Georgii},\ and\ \citenamefont {Böni}}]{Muehlbauer:09b}%
  \BibitemOpen
  \bibfield  {author} {\bibinfo {author} {\bibfnamefont {S.}~\bibnamefont
  {Mühlbauer}}, \bibinfo {author} {\bibfnamefont {B.}~\bibnamefont {Binz}},
  \bibinfo {author} {\bibfnamefont {F.}~\bibnamefont {Jonietz}}, \bibinfo
  {author} {\bibfnamefont {C.}~\bibnamefont {Pfleiderer}}, \bibinfo {author}
  {\bibfnamefont {A.}~\bibnamefont {Rosch}}, \bibinfo {author} {\bibfnamefont
  {A.}~\bibnamefont {Neubauer}}, \bibinfo {author} {\bibfnamefont
  {R.}~\bibnamefont {Georgii}}, \ and\ \bibinfo {author} {\bibfnamefont
  {P.}~\bibnamefont {Böni}},\ }\href@noop {} {\bibfield  {journal} {\bibinfo
  {journal} {Science},\ }\textbf {\bibinfo {volume} {323}},\ \bibinfo {pages}
  {915} (\bibinfo {year} {2009})}\BibitemShut {NoStop}%
\end{thebibliography}
\end{document}